\documentclass[final]{oscmjournal}
\usepackage[utf8]{inputenc}
\usepackage{graphicx}
\usepackage{hyperref}
\usepackage{xcolor}
\usepackage{csquotes}
\usepackage{amssymb}
\usepackage{amsmath}
\usepackage{multirow}
\usepackage{booktabs}
\usepackage{array}
\usepackage{float}

\usepackage{array}
\newcolumntype{P}[1]{>{\raggedright\arraybackslash}p{#1}}

\title{An Efficient Intelligent Semi-Automated Warehouse Inventory Stocktaking System}
\authorinfo{
    \textbf{Chunan Tong, CSCP-F}\\    
    University of Maryland\\
    Email: tcn1989@terpmail.umd.edu\\
}

\begin{document}

\maketitle

\begin{abstract}
As supply chain management continues to evolve, efficient inventory management systems have become increasingly crucial. However, traditional manual methods often struggle to meet the complexities of modern market demands, particularly when it comes to data accuracy, delays in monitoring, and the heavy reliance on subjective experience for forecasting. This study introduces an intelligent, semi-automated barcode-based inventory management system designed to overcome these challenges. The system integrates barcode technology with a distributed architecture, combined with big data analytics and machine learning for real-time tracking and accurate inventory predictions. Its performance has been validated through multiple simulation tests, where it has outperformed traditional RFID technology in certain cases. Through careful system design, technology exploration, and validation, this research demonstrates the significant potential of this intelligent system in improving inventory management efficiency and accuracy.
\end{abstract}

\section{INTRODUCTION}\label{sec:introduction}
In recent years, fully automated inventory technologies have experienced significant advancements, with the utilization of next-generation ultra-high-frequency RFID (Radio-Frequency Identification) employing radio waves for real-time identification and tracking of tagged objects. The application of RFID in warehouse management traces back to the 1990s, with initiatives by major players such as Walmart and the US Department of Defense. However, it wasn't until the 2000s that RFID technology gained widespread adoption due to the establishment of standards and cost reductions \parencite{wyld2006rfid}. While new RFID systems have demonstrated impressive accuracy in real-time inventory monitoring and complex tasks like expiration date tracking, the full extent of their potential remains unpredictable in research of \parencite{ritter1951stocktaking}. This raises questions about the limited adoption of these powerful systems in large manufacturing factories.

Although fully automated inventory technology is promising, warehouse management is not yet ready for full adoption. Prior research, such as \parencite{dong2008load}, has highlighted a fundamental issue in large-scale RFID deployments: the complexity of load balancing in dense inventory environments. Their analysis demonstrated that minimizing RFID reader overload is an NP-hard problem, meaning that maintaining consistent performance in high-density warehouses is computationally intensive and inherently inefficient. This challenge is further exacerbated by environmental factors—\parencite{zhu2007solutions} found that RFID accuracy significantly deteriorates when tags are placed near metals or liquids, making RFID unreliable in diverse warehouse conditions.

Beyond technical limitations, RFID adoption also faces economic and operational barriers. \parencite{huber2007barriers} identified that the high cost of RFID tags, readers, and integration with existing warehouse management systems creates a financial burden for businesses. While RFID promises real-time stock tracking, achieving full automation often requires substantial infrastructure investment, which many companies find prohibitive. Furthermore, studies such as \parencite{white2007comparison} indicate that fully automated RFID inventory systems may not be as cost-effective as initially anticipated, due to ongoing maintenance expenses and frequent calibration requirements.

Due to these limitations, it is clear that a fully automated RFID-based inventory management system is not always the optimal solution, particularly for warehouses transitioning from traditional manual stocktaking methods. The gap between RFID’s potential and its real-world feasibility necessitates an alternative approach that balances automation with operational practicality. This study proposes a semi-automated inventory system that leverages barcode scanning and real-time data analytics to bridge this gap. Unlike RFID, which suffers from environmental interference and high implementation costs, barcode-based inventory tracking provides a more stable, cost-effective, and adaptable alternative.

The proposed \textbf{semi-automated inventory system} integrates handheld barcode scanning terminals with an intelligent backend architecture, ensuring real-time ERP synchronization and dynamic stock management. The system follows a client-server model, where Android-based PDA devices capture barcode data and communicate with the backend via a RESTful API. The backend, running on an Ubuntu Linux environment with a distributed Hadoop HDFS database, cross-checks scanned records against SAP S/4HANA ERP for real-time stock validation.

A key innovation of this system is integrating Apache Flink’s watermarking mechanism, which enables real-time tracking of inbound and outbound inventory. By leveraging event-time processing, Flink dynamically updates inventory levels, ensuring high accuracy even in high-throughput warehouse environments. This approach effectively mitigates latency issues in traditional stock reconciliation, allowing instant discrepancy detection.

To improve anomaly detection and stock integrity, the backend uses AI-driven analytics with the Isolation Forest algorithm. It converts Bin, Batch, and Handling Unit (HU) data into feature vectors and assigns anomaly scores to identify discrepancies.

This hierarchical anomaly detection method fully utilizes structured stocktaking data to improve inventory accuracy, enable real-time monitoring, and prevent errors proactively. Combined with scalable big data processing, it provides a cost-effective, flexible alternative to RFID-based solutions, making it highly suitable for modern warehouse management and digital transformation.

By critically assessing RFID’s limitations and offering a structured alternative, this research provides a clear pathway for businesses seeking to modernize their inventory management without the financial and operational burdens associated with full RFID adoption. This study contributes to the field by presenting a scalable and pragmatic inventory solution, offering valuable insights for warehouse managers and supply chain professionals navigating the transition toward digitalized stock control \parencite{sciullo2023design}.

\section{LITERATURE REVIEW}\label{sec:literature_review}
Barcode-based inventory management systems have long been a staple in warehouse operations due to their affordability and simplicity in \parencite{rossetti2001inventory}. However, despite their widespread use, they exhibit several limitations, especially in large-scale or high-turnover environments. One major issue with traditional barcode systems is the reliance on batch synchronization for inventory updates. As a result, data is often outdated by the time it is processed, which hinders timely decision-making. The manual scanning process, though essential for the accuracy of the data, introduces delays as updates occur only once the scanning process is complete.

The dependence on human labor also opens the door to various types of errors. Barcode scanning systems are still heavily reliant on operators manually scanning each item, which leads to mistakes such as missed scans, misidentification of items, or duplication of scans. Such errors accumulate over time, significantly affecting inventory accuracy and requiring additional manual checks to resolve discrepancies in \parencite{ngai2008rfid}. As warehouses grow, these inefficiencies become more pronounced, limiting the scalability and operational effectiveness of barcode-based systems.

Existing barcode systems also lack intelligent anomaly detection mechanisms. Discrepancies in inventory data are often not identified until manually flagged by operators, which can delay responses to issues like stockouts or overstocking. This lack of real-time anomaly detection exacerbates operational inefficiencies, particularly in large-scale warehouses where the time taken to identify and correct discrepancies is critical.

\subsection{Semi-Automated Barcode-Based Systems}
Semi-automated barcode-based systems, while still reliant on manual scanning, have been increasingly integrated with software systems that streamline the inventory process. Research has shown that integrating barcode scanning with cloud-based platforms can help improve data accuracy and enable real-time synchronization between the warehouse and ERP systems. A study by \parencite{muyumba2017web} addresses the inventory management issues of the Zambia Air Force (ZAF) by proposing an automated inventory management system based on cloud architecture and barcode technology to resolve problems caused by manual operations, such as inventory errors, inefficient management, item losses, and high costs. System testing demonstrated that barcode scanning improved inventory accuracy by 95\%, was over three times faster than manual management, and enhanced security through CCTV monitoring. However, the study has certain limitations, including the lack of consideration for more advanced technologies such as RFID, insufficient security analysis, unquantified CCTV effectiveness, and the absence of a return on investment (ROI) assessment.

\subsection{RFID-Based Semi-Automated System}
Similarly, RFID represents the latest technology in next-generation warehouse inventory management and has been included in comparative studies to evaluate its advantages and limitations in improving inventory management efficiency. A study by \parencite{white2007comparison} conducted an experimental comparison between barcoding and radio frequency identification (RFID) technologies in inventory management, focusing on scanning speed, accuracy, equipment failure rate, and applicable environments, with testing carried out in a cold chain warehouse. The results showed that RFID scanning speed was 2.5 times faster than barcoding, but its error rate (46.5\%) and equipment failure rate (37\%) were higher, and its performance in cold chain environments was unstable. The study also highlighted that two-dimensional barcodes (2D Barcode) are narrowing the performance gap with RFID and suggested that future research should explore hybrid systems that integrate the advantages of both technologies. However, the study lacked detailed cost analysis and long-term return on investment (ROI) evaluation and was limited to cold chain warehouses, without addressing the needs of other industries. Future advancements in expanding the scope of experiments, optimizing RFID equipment, researching hybrid systems, and analyzing its potential in supply chain big data management could enhance the intelligence and efficiency of inventory management. However, while these solutions address some of the shortcomings of traditional barcode systems, they still rely heavily on manual scanning and lack intelligent anomaly detection capabilities, leaving them vulnerable to human error and delayed issue resolution.

\parencite{dong2008load} revealed that the load balancing problem in large-scale RFID systems shows that RFID readers in high-density tag environments are prone to uneven load distribution, data loss, and excessive energy consumption. However, the study does not explore dynamic environments, cost analysis, and security issues in depth, which share certain similarities with the conclusions drawn by \parencite{white2007comparison}, which also served as one of the motivations for this research.

RFID technology consists of three main components:
\begin{itemize}
    \item \textbf{RFID Tags:} embedded microchips attached to inventory items for unique identification,
    \item \textbf{RFID Readers:} devices that emit radio waves to activate and read RFID tags, and
    \item \textbf{RFID Middleware:} a software layer responsible for processing tag data and integrating with warehouse management systems (WMS) or enterprise resource planning (ERP) platforms.
\end{itemize}
This technology sometimes encounter several challenges, including:
\begin{enumerate}
    \item \textbf{Readability Issues in Complex Environments} \\
    Signal interference from metal surfaces, liquids, and electromagnetic interference (EMI) significantly affects scanning accuracy.  
    The probability of a successful RFID read \( P_{\text{read}} \) decreases with increasing tag density:
    \begin{equation}
    P_{\text{read}} = 
    \begin{cases}
    1 - \frac{N_{\text{tags}}}{N_{\text{max}}}, & \text{if } N_{\text{tags}} \leq N_{\text{max}}, \\
    0, & \text{if } N_{\text{tags}} > N_{\text{max}},
    \end{cases}
    \end{equation}
    where \( N_{\text{tags}} \) is the number of RFID tags in the scanning zone, and \( N_{\text{max}} \) is the reader's tag capacity before data loss occurs.
    \item \textbf{High Cost of Implementation} \\
    RFID deployment involves substantial infrastructure investment, including RFID readers, middleware, and ERP integration.  
    Tagging costs remain high, especially for environments requiring metal-compatible RFID tags.
    \item \textbf{Network Congestion and Data Latency} \\
    Large-scale RFID deployments produce high-frequency data transmissions, causing network congestion and data loss:
    \begin{equation}
    P_{\text{loss}} = 1 - \frac{C_{\text{max}}}{M \times R},
    \end{equation}
    where \( C_{\text{max}} \) is server capacity, \( M \) is the number of RFID readers, and \( R \) is the data rate per reader.
    \item \textbf{Complex Integration with Existing Systems} \\
    Unlike barcode-based solutions, RFID requires specialized ERP connectors for real-time processing.  
    Many legacy warehouse management systems lack native RFID support, necessitating additional middleware.
\end{enumerate}
These challenges necessitate an alternative inventory tracking approach that balances automation, cost-effectiveness, and ease of deployment.

\subsection{Cutting-Edge Technologies in Stocktaking}
\subsubsection{IoT-Based Inventory Tracking System}
The Internet of Things (IoT) represents a transformative approach to real-time inventory management by enabling the interconnection of physical assets through wireless networks, sensors, and cloud-based platforms. IoT systems integrate various components such as RFID, BLE, GPS, LPWAN (LoRa, NB-IoT, LTE-M), and smart devices to collect, process, and transmit inventory-related data \parencite{mashayekhy2022impact}.

\begin{table*}[htbp]
\sffamily \small
\centering
\caption{Comparison of LoRa, NB-IoT, and LTE-M in Inventory Auditing}
\label{tab:lpwan-comparison}
\resizebox{\textwidth}{!}{%
\scriptsize
\begin{tabular}{P{1.5cm}P{3.5cm}P{2.3cm}P{3cm}P{3cm}}
\toprule
\textbf{Technology} & \textbf{Frequency Band and Power Consumption} & \textbf{Coverage Range} & \textbf{Data Rate and Latency} & \textbf{Applications} \\
\midrule
\textbf{LoRa} \text{(LoRaWAN)} & Unlicensed spectrum, ultra-low power (1 mAh to join; 44-byte uplink: 100 $\mu$Ah) & Up to 20 km rural, 5~km urban & Low rate (up to 50 kbps); high latency (2--3.5 sec) & Static inventory tracking (e.g., fixed assets in warehouses)\textsuperscript{1} \\
\addlinespace[0.8ex]
\textbf{NB-IoT} & Licensed LTE spectrum, moderate power (3 mAh to join; 44-byte uplink: 1.8 mAh) & $\sim$4 km rural, 1~km~\text{urban} & Higher throughput (up to 200 kbps); low latency ($\sim$244 ms) & Dynamic inventory tracking (e.g., shelf monitoring in retail)\textsuperscript{2} \\
\addlinespace[0.8ex]
\textbf{LTE-M} & Licensed spectrum, higher power; supports mobility & Similar to NB-IoT; optimized for mobility & Higher data rate (up to 200 kbps); lower latency & Mobile inventory tracking (e.g., goods-in-transit)\textsuperscript{3} \\
\bottomrule
\end{tabular}%
}
\vspace{1ex}
\begin{minipage}{\textwidth}
\textsuperscript{1}Semtech used LoRaWAN for airport baggage tag tracking, reducing costs by 30--40\%. \newline
\textsuperscript{2}Link Labs demonstrated reliable NB-IoT applications in warehouse asset tracking. \newline
\textsuperscript{3}Widely used in asset tracking with high update frequency and data rate requirements.
\end{minipage}
\end{table*}

Among the enabling technologies, LPWAN solutions such as LoRa, NB-IoT, and LTE-M are particularly suited for large-scale inventory tracking due to their long-range connectivity and low power consumption. Complementary technologies like RFID and BLE also play vital roles: RFID excels in short-range, high-precision tracking (e.g., shelf management in warehouses), while BLE supports low-power indoor positioning (e.g., retail asset monitoring). For outdoor scenarios, GPS is often paired with LPWAN to track mobile inventory, such as goods in transit. Table~\ref{tab:lpwan-comparison} compares the technical characteristics and applications of LoRa, NB-IoT, and LTE-M in inventory auditing.

In sum, the effectiveness of IoT technologies in the market depends on factors such as the distance from the base station and signal density, similar to mobile phones \parencite{mugerwa2024adaptive}. In many cases, warehouses are situated in enclosed environments, which can lead to poor signal reception, hindering IoT devices from transmitting their heartbeat packets back to the server \parencite{morillo2024technology, sciullo2023design}. Additionally, while advanced LoRa IoT nodes can communicate with each other, this inter-node communication may cause network congestion in both uplink and downlink channels, potentially leading to system failures \parencite{sciullo2023design}. However, a significant advantage of IoT devices is their low power consumption; they often send heartbeat packets infrequently, allowing batteries to last over five years \parencite{farhad2023mobility}.

\subsubsection{Inventory Tracking System Using Drone}
Drone-assisted inventory systems, leveraging Unmanned Aerial Vehicles (UAVs), represent a transformative approach to automated warehouse management tasks such as stocktaking and item tracking. These systems are particularly advantageous in large-scale facilities. The primary applications include:
\begin{itemize}
    \item \textbf{Automated Inventory Counting}: Drones equipped with barcode scanners or RFID readers can cruise through the warehouse to identify and record inventory data, significantly reducing manual labor and time. For example, drones are capable of scanning hundreds of locations per hour.
    \item \textbf{Cyclical Stocktaking}: Drones can perform regular inventory audits without disrupting day-to-day operations.
    \item \textbf{Enhanced Safety}: By accessing high or narrow areas, drones reduce the risks associated with manual handling.
    \item \textbf{Real-time Updates and Management}: These systems provide real-time inventory data, especially useful for managing perishable items using the First-Expire-First-Out (FEFO) strategy to improve efficiency \parencite{singh2024drone}.
\end{itemize}

Current evidence shows that drones excel in inspecting elevated warehouse areas. However, challenges remain in terms of cost, safety, regulatory compliance, and compatibility with warehouse layout \parencite{kumar2023drone}. Therefore, although drone-assisted inventory systems hold great potential in warehouse automation, their effectiveness depends on overcoming technological, economic, and operational barriers, particularly in developing solutions for scanning stacked inventory.

Despite the promise of drone-assisted inventory systems, their widespread application is hindered by several limitations, summarized in Table~\ref{tab:drone-limitations}. Moreover, the integration of Artificial Intelligence (AI) and Machine Learning (ML) has significantly enhanced the accuracy of label scanning, reaching precision levels of 97\% to 99\% \parencite{morillo2024technology}, which constitutes an unexpected but noteworthy advantage.

\begin{table*}[htbp]
\sffamily \small
\centering
\caption{Limitations of Drone-Assisted Inventory Systems}
\label{tab:drone-limitations}
\footnotesize
\resizebox{\textwidth}{!}{%
\begin{tabular*}{\textwidth}{@{\extracolsep{\fill}}P{4cm}P{13cm}}
\toprule
\textbf{Limitation} & \textbf{Description} \\
\midrule
Battery Life Constraints & Limited battery life requires efficient path planning or the deployment of multiple drones to cover large areas. \\
Indoor Navigation & The absence of GPS signals indoors necessitates the use of costly SLAM (Simultaneous Localization and Mapping) or beacon systems for navigation. \\
Stacked Inventory Limitation & In densely stored or vertically stacked environments, drones struggle to scan labels obscured by other items, leading to decreased accuracy. \\
Operational Safety & Robust obstacle avoidance mechanisms are essential to ensure safe operation. \\
High Initial Cost & The cost of hardware, software, and integration with Warehouse Management Systems (WMS) may be prohibitive for small businesses. \\
\bottomrule
\end{tabular*}
}
\normalsize\normalfont
\end{table*}

\subsection{State-of-the-Art Comparison}
In the realm of supply chain optimization, selecting an effective inventory management system hinges on a systematic evaluation of key criteria and their assigned weights. \parencite{chopra2016supply} underscore the importance of inventory accuracy and real-time data, while \parencite{lee2004information} stress the critical role of information accuracy and timeliness. Drawing from these foundational works, we identified five essential evaluation criteria: cost efficiency, inventory accuracy, implementation complexity, real-time performance, and scalability. The Analytic Hierarchy Process (AHP) was utilized to determine their relative importance through expert pairwise comparisons.

\subsubsection{AHP Derivation Process}
AHP employs a 1-9 scale to quantify expert judgments, where 1 signifies equal importance and 9 denotes extreme importance. Experts in this study prioritized \textit{cost efficiency} and \textit{inventory accuracy} as key drivers in reducing supply chain inefficiencies. The weight is calculated as follows:

\begin{enumerate}
    \item \textbf{Matrix Normalization}: Each column is normalized by dividing its elements by the column sum. For example, the first column sums to 
    \begin{equation*}
    1 + \frac{1}{3} + \frac{1}{5} + \frac{1}{4} + \frac{1}{4} = 2.033,
    \end{equation*}
    yielding normalized values such as $1/2.033= 0.492$, $(1/3)/2.033 = 0.164$, etc.
    \item \textbf{Eigenvector Calculation}: Weights are computed by averaging the rows of the normalized matrix, yielding cost efficiency 47.6\%, inventory accuracy 21.0\%, implementation complexity 7.1\%, real-time performance 12.1\%, and scalability 12.1\%.
    \item \textbf{Consistency Check}: The consistency ratio (CR) of 0.045, below the 0.1 threshold, confirms the matrix's reliability.
\end{enumerate}
The resulting judgment matrix is shown in Table 3.

\begin{table*}[htbp]
\sffamily \small
\centering
\caption{Judgment Matrix for Evaluation Criteria}
\resizebox{\textwidth}{!}{ % Scales the table to fit within \textwidth
\begin{tabular}{@{}l|ccccc@{}} % Removed \extracolsep to avoid unnecessary stretching
\toprule
& Cost Efficiency & Inventory Accuracy & Implementation Complexity & Real-Time Performance & Scalability \\
\midrule
Cost Efficiency & 1 & 3 & 5 & 4 & 4 \\
Inventory Accuracy & 1/3 & 1 & 3 & 2 & 2 \\
Implementation Complexity & 1/5 & 1/3 & 1 & 1/2 & 1/2 \\
Real-Time Performance & 1/4 & 1/2 & 2 & 1 & 1 \\
Scalability & 1/4 & 1/2 & 2 & 1 & 1 \\
\bottomrule
\end{tabular}
} % End \resizebox
\normalsize\normalfont
\end{table*}

\subsubsection{System Evaluation Results}
Three inventory management systems were assessed using these weights on a 10-point scale, with results presented in Table \ref{tab:sys_eval}. The Barcode-Flink-AI system emerged as the top performer, excelling in cost efficiency and inventory accuracy—criteria with the highest weights.

\begin{table}[htbp]
\sffamily \small
\centering
\caption{System Evaluation Scores}
\label{tab:sys_eval}
\footnotesize
\begin{tabular*}{\columnwidth}{@{\extracolsep{\fill}}p{4cm}c}
\toprule
\textbf{System} & \textbf{Weighted Score} \\
\midrule
Barcode-Flink-AI System & 9.15 \\
IoT-Based System & 5.6 \\
Drone-Assisted System & 4.3 \\
\bottomrule
\end{tabular*}
\normalsize\normalfont
\end{table}

\subsubsection{Robustness Analysis Under Weight Variations}
A sensitivity analysis was conducted to evaluate the Barcode-Flink-AI system's adaptability by adjusting weight distributions to reflect varying supply chain priorities, such as raising real-time performance to 30\% and lowering cost efficiency to 35\% or increasing scalability to 25\% while reducing inventory accuracy to 15\%. Across these scenarios, the Barcode-Flink-AI system's score ranged from 8.8 to 9.3, consistently outperforming competitors. This resilience highlights its robustness and versatility.

\subsubsection{Comparative Analysis Across Key Dimensions}
Table \ref{tab:system-comparison} compares the Barcode-Flink-AI system with IoT-based and drone-assisted systems across critical dimensions, emphasizing its advantages in cost, integration, and adaptability.

\begin{table*}[htbp]
\sffamily \small
\centering
\caption{Comparison of Inventory Management Systems Across Key Dimensions}
\label{tab:system-comparison}
\footnotesize
\begin{tabular*}{\textwidth}{@{\extracolsep{\fill}}p{3.5cm}p{4cm}p{4.5cm}p{4cm}}
\toprule
\textbf{Dimension} & \textbf{Barcode-Flink-AI System} & \textbf{IoT-Based System} & \textbf{Drone-Assisted System} \\
\midrule
\textbf{Cost} & Low ($\sim$18\% of RFID costs) & High (sensors, cloud infrastructure) & High (hardware, navigation) \\
\addlinespace[0.5ex]
\textbf{Integration} & Seamless with WMS/ERP & Complex; network-dependent & Requires specialized navigation \\
\addlinespace[0.5ex]
\textbf{Real-Time Performance} & Near real-time (99.2\% accuracy) & Real-time; network-reliant & Real-time; battery-limited \\
\addlinespace[0.5ex]
\textbf{Ease of Use} & Minimal training required & Technical expertise needed & Drone skills required \\
\addlinespace[0.5ex]
\textbf{Adaptability} & Flexible across layouts & Suited for large operations & Limited to drone-friendly setups \\
\addlinespace[0.5ex]
\textbf{Accuracy} & Maximum 99.2\% with AI detection & High with proper setup & 97\%-99\% with AI \\
\bottomrule
\end{tabular*}
\normalsize\normalfont
\end{table*}

\subsection{Research Gap}
Current research in semi-automated inventory management often focuses on RFID as the primary solution. However, there is a growing need for alternative approaches that are cost-effective, scalable, and easy to deploy. Existing studies on barcode-based systems have explored their integration with cloud computing and mobile devices \parencite{muyumba2017web}, but these systems still fall short in addressing the need for real-time updates and intelligent anomaly detection.

To fill this gap, our system combines the best of barcode technology with cutting-edge stream processing and AI capabilities, offering a more balanced solution that minimizes costs, reduces errors, and scales efficiently. We compare our system to RFID-based, traditional manual inventory counting, and barcode-based systems in Table~\ref{tab:system_comparison}.

\begin{table*}[htbp]
\sffamily \small
\centering
\caption{Comparison of Inventory Management Approaches}
\label{tab:system_comparison}
\footnotesize
\begin{tabular*}{\textwidth}{@{\extracolsep{\fill}}P{0.2\textwidth}p{0.23\textwidth}p{0.22\textwidth}p{0.25\textwidth}}
\toprule
\textbf{Feature} & \textbf{RFID-Based System} & \textbf{Manual Stocktaking} & \textbf{Proposal (Flutter + Flink + AI)} \\
\midrule
\textbf{Cost} & High (\$500K+ for full deployment) & Low (Minimal investment) & Low (Standard barcode scanners + PDA) \\
\textbf{Stocktaking Speed} & Fast (affected by signal interference) & Slow (Manual item-by-item scanning) & Fast (Automated barcode scanning + AI anomaly detection) \\
\textbf{Error Rate} & Moderate (RFID misreads, environmental factors) & High (Human errors) & Low (AI-assisted detection, automated validation) \\
\textbf{Real-time Sync} & Requires additional ERP integration & No real-time synchronization & Instant updates via Flink stream processing \\
\textbf{Scalability} & Requires expensive infrastructure & Limited (Depends on workforce availability) & High (Easily adaptable) \\
\textbf{Environmental Reliability} & Prone to interference (metal/liquids) & Not affected & Not affected (Barcode scanning is robust) \\
\textbf{Ease of Use} & Requires training & Time-consuming & User-friendly with AI-assisted alerts \\
\textbf{Operational Efficiency} & Medium (some automation, but requires ERP adjustments) & Low (labor-intensive, slow updates) & High (AI minimizes manual intervention, increases efficiency) \\
\textbf{Maintenance Cost} & High (frequent tag replacements, system tuning) & Low (minimal maintenance) & Low (barcode-based, minimal upkeep) \\
\textbf{Implementation Complexity} & High (complex hardware/software integration) & Low (no complex infrastructure required) & Low (Standard barcode scanning + lightweight cloud integration) \\
\textbf{Training Time for Operators} & 6+ hours (Needs training in RFID usage and ERP) & 3+ hours (Manual stocktaking procedures) & 1-2 hours (Simple barcode scanning with AI-assisted alerts) \\
\textbf{Overall Cost-Effectiveness} & Expensive (High initial cost, maintenance, and operational expenses) & Low-cost (But inefficient in large-scale operations) & Optimal balance of cost, speed, and accuracy \\
\bottomrule
\end{tabular*}
\normalsize\normalfont
\end{table*}

\section{METHODOLOGY}\label{sec:methodology}
\begin{figure*}[htbp]
    \centering
    \includegraphics[width=1\textwidth]{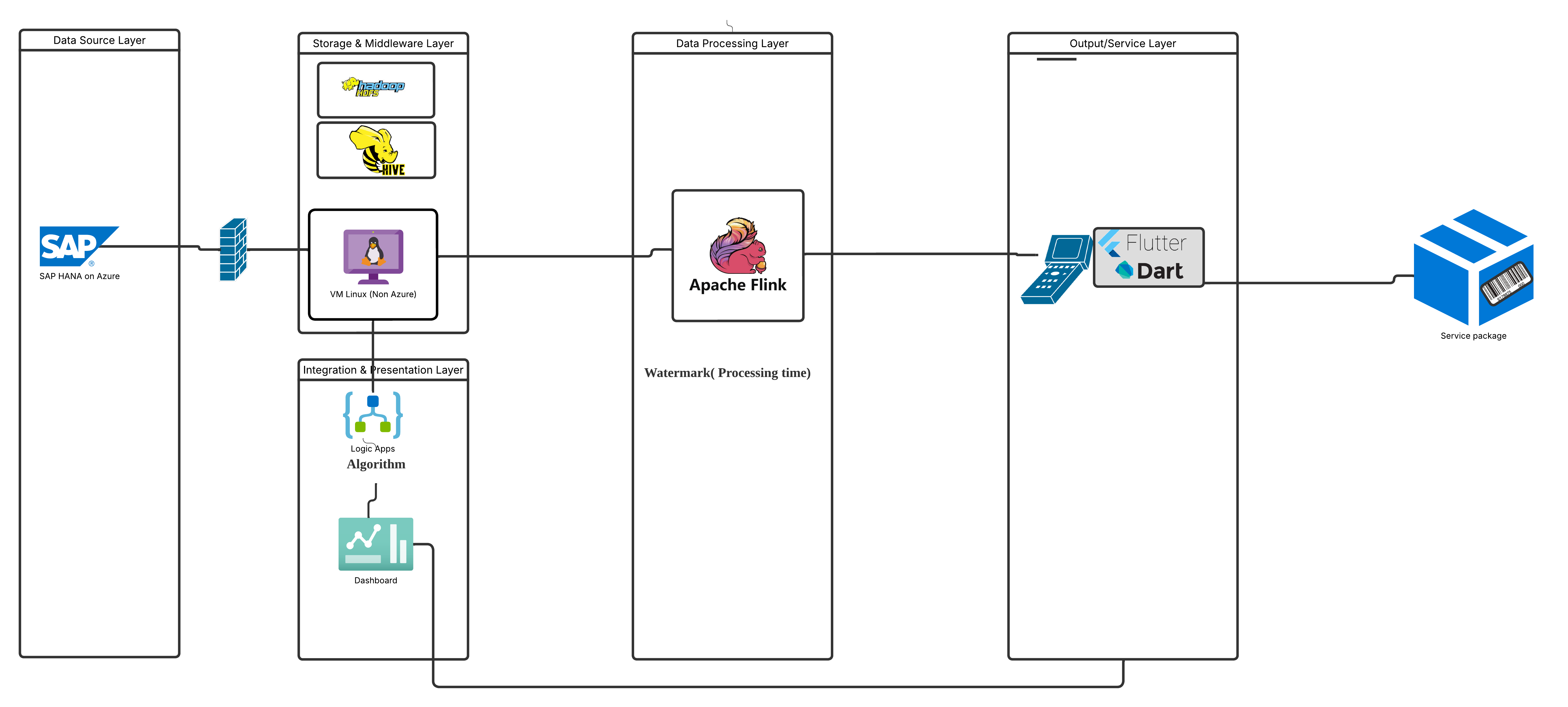}
    \caption{System Architecture Diagram}
    \label{fig:enter-label}
\end{figure*}

Figure \ref{fig:enter-label} illustrates the overall architecture of the developed system. The system architecture consists of four core components. First, handheld PDA terminals running a Flutter-based inventory application interact with the backend through a REST API. Second, a distributed processing backend utilizes Apache Flink for event-time processing, ensuring continuous stock updates. Third, anomaly detection and analytics are performed using Isolation Forest algorithms, which detect stock inconsistencies in real time. Finally, cloud-based ERP integration enables direct synchronization with SAP S/4HANA via WebSocket, ensuring dynamic inventory validation.

This section describes the core architecture of our proposed semi-automated inventory system. The system is designed to provide real-time inventory updates, AI-driven anomaly detection, and seamless integration with enterprise resource planning (ERP) systems. The three main components of the system include a Flutter-based mobile terminal, real-time data processing using Apache Flink, and AI-powered anomaly detection with the Isolation Forest algorithm.

\subsection{Anomaly Detection with Isolation Forest}
The proposed system utilizes the Isolation Forest for anomaly detection, an algorithm well-suited for identifying outliers in high-dimensional datasets. However, other alternatives, such as DBSCAN and AutoEncoder, were considered for comparison, though, the algorithm comes with a few advantages compared to these alternatives. First, it has low computational complexity and is scalable for large-scale inventory data. It also operates in an unsupervised manner, requiring no labeled anomaly data. In addition, it is robust in handling inventory with varying SKU distributions, and, compared to DBSCAN and AutoEncoders, it provides stable results without requiring extensive data preprocessing or labeled anomaly samples. The comparison is summarized in Table~\ref{tab:anomaly_detection_comp}.

\begin{table*}[htbp]
\sffamily \small
\centering
\caption{Comparison of Anomaly Detection Methods}
\label{tab:anomaly_detection_comp}
\footnotesize
\begin{tabular*}{\textwidth}{@{\extracolsep{\fill}}p{4cm}p{5cm}p{5cm}}
\toprule
\textbf{Method} & \textbf{Advantages} & \textbf{Disadvantages} \\
\midrule
\textbf{Isolation Forest (IF)} & Efficient for high-dimensional data; Computationally efficient (\( O(n \log n) \)) & Sensitive to parameter tuning; Not optimal for time-series data \\
\textbf{DBSCAN} & Detects density-based anomalies; Robust to noise & Requires manual parameter tuning (\( \varepsilon \)); Poor performance in high dimensions \\
\textbf{AutoEncoder (AE)} & Learns complex inventory patterns; Suitable for sequential data & High computational cost; Requires large training datasets \\
\bottomrule
\end{tabular*}
\normalsize\normalfont
\end{table*}

\subsubsection{Data Preprocessing}
Let us assume we have a dataset containing multiple handling units (HUs), where each HU contains: product quantity \( x_{hu} \), batch \( b_{hu} \), and bin location \( bin_{hu} \). We organize the data as follows:
\begin{equation}
\mathcal{D} = \{ (x_{hu}, b_{hu}, bin_{hu}) \}, \quad hu \in \{1, 2, \dots, N_{hu}\},    
\end{equation}
where \( N_{hu} \) is the total number of handling units. The data of each HU is used for training and detecting anomalies.

\subsubsection{Isolation Forest (IF) Model}
IF is a tree-based anomaly detection method typically used for high-dimensional data. In our approach, we use the algorithm to detect anomalies in the product quantity of each HU. Suppose we have trained an IF model, and the model’s goal is to calculate anomaly scores based on the “isolation degree” of the sample. The anomaly score \( s(x_{hu}) \) is calculated as:
\begin{equation}
s(x_{hu}) = \frac{1}{T} \sum_{t=1}^{T} h_t(x_{hu}), 
\end{equation}
where \( T \) is the total number of trees, and \( h_t(x_{hu}) \) is the score from the \( t \)-th tree for sample \( x_{hu} \). A lower score indicates a higher likelihood of the sample being an anomaly.

\subsubsection{Batch and Bin Level Data Aggregation}
In addition to the HU-level data, we also aggregate data at the BATCH and BIN levels to improve the accuracy of anomaly detection. Specifically, we compute the following statistics at the BATCH and BIN levels:
\begin{itemize}
    \item BATCH-level mean product quantity \( \mu_b \):
    \begin{equation}
        \mu_{b} = \frac{1}{N_{b}} \sum_{hu \in B_{b}} x_{hu},
    \end{equation}    
    where \( N_{b} \) is the number of HUs in batch \( b \), and \( B_b \) is the set of HUs belonging to batch \( b \).
    \item BATCH-level standard deviation \( \sigma_b \):
    \begin{equation}
        \sigma_{b} = \sqrt{\frac{1}{N_{b}} \sum_{hu \in B_{b}} (x_{hu} - \mu_b)^2}.
    \end{equation}
    \item BIN-level mean product quantity \( \mu_{bin} \):
    \begin{equation}
        \mu_{bin} = \frac{1}{N_{bin}} \sum_{hu \in BIN_{bin}} x_{hu},
    \end{equation}
    where \( N_{bin} \) is the number of HUs in bin \( bin \), and \( BIN_{bin} \) is the set of HUs belonging to bin \( bin \).
\end{itemize}

\subsubsection{Data Fusion and Joint Feature Construction}
To fully leverage the data at each level, we construct joint feature vectors that combine the information from HU, BATCH, and BIN. Specifically, we construct a feature vector for each HU that includes the following features:
\begin{equation}
    \mathbf{f}_{hu} = \left( x_{hu}, \mu_{b_{hu}}, \sigma_{b_{hu}}, \mu_{bin_{hu}}, \sigma_{bin_{hu}} \right),
\end{equation}
where \( x_{hu} \) is the product quantity of the HU, \( \mu_{b_{hu}} \) and \( \sigma_{b_{hu}} \) are the mean and standard deviation of the batch that the HU belongs to, and \( \mu_{bin_{hu}} \) and \( \sigma_{bin_{hu}} \) are the mean and standard deviation of the bin that the HU belongs to. This feature vector not only captures the product quantity information of the individual HU but also considers global statistical information from its associated BATCH and BIN.

\subsubsection{Combined Anomaly Score Calculation}
For each HU, we first compute the anomaly score based on the IF model
\begin{equation}
    s(x_{hu}) = \frac{1}{T} \sum_{t=1}^{T} h_t(x_{hu}).
\end{equation}
We then compute a weighted combined anomaly score by considering the statistics from the different levels. The combined anomaly score \( S_{total}(x_{hu}) \) is calculated as:
\begin{align}
    S_{total}(x_{hu}) &= \alpha \cdot s(x_{hu}) + \beta \cdot \left( \frac{| \mu_{b_{hu}} - x_{hu} |}{\sigma_{b_{hu}}} \right) \nonumber \\
    &+ \gamma \cdot \left( \frac{| \mu_{bin_{hu}} - x_{hu} |}{\sigma_{bin_{hu}}} \right),
\end{align}
where \( \alpha, \beta, \gamma \) are the weighting coefficients that adjust the influence of each feature. \( s(x_{hu}) \) is the anomaly score based on Isolation Forest, and \( \mu_{b_{hu}}, \sigma_{b_{hu}} \) are the mean and standard deviation of the batch that the HU belongs to, and \( \mu_{bin_{hu}}, \sigma_{bin_{hu}} \) are the mean and standard deviation of the bin that the HU belongs to.

\subsubsection{Distributed Computation and Large-Scale Processing}
Since our system needs to handle large volumes of real-time data, we use distributed computing methods and leverage frameworks such as Apache Spark for efficient parallel computation. Specifically, we partition the data into multiple chunks and use the distributed computing framework to calculate anomaly scores for each HU, as well as the statistics for BATCH and BIN levels. Finally, we perform distributed aggregation to compute the final anomaly score for each HU and determine the root causes of anomalies.

In Spark, we use the following operations:
\begin{itemize}
    \item \textbf{Data partitioning and mapping}: Partition the data across different compute nodes.
    \item \textbf{Parallel computation}: Each node computes the anomaly score for individual HUs and the statistics for Batch and Bin levels.
    \item \textbf{Result aggregation and merging}: Aggregate the computed results from different nodes to calculate the final anomaly scores.
\end{itemize}

\subsubsection{Anomaly Detection and Real-Time Response}
When an anomaly is detected, our system triggers a real-time response mechanism that promptly notifies operators to inspect the data. The anomaly detection results not only identify which HUs have issues but also pinpoint the potential root cause (e.g., a problematic BATCH or BIN). For example, if an anomaly is detected in a particular BATCH due to a large fluctuation in product quantities, the system will alert the operator to inspect all HUs within that batch.

\subsection{Event Ordering with Watermark}
In this project, we adopted Apache Flink's Watermark mechanism to handle real-time inventory data streams. The Watermark mechanism ensures that events are processed based on event time, which is crucial for accurately managing the sequence of inventory events in real time. Mathematically, the Watermark process can be described as follows:
\begin{equation}
W(t) = \max(\text{Event Time}) - \Delta t,    
\end{equation}
where \( W(t) \) is the watermark at time \( t \), and \( \Delta t \) is the maximum allowed lateness for late-arriving events. This formula means that the watermark is advanced based on the maximum event time observed up to that point, but no event older than \( \Delta t \) will be processed. This ensures that events are processed in their correct event time order, even if they arrive out of order.

\subsubsection{Late Data Issues in Inventory Reconciliation}
Late-arriving data is a common challenge in inventory management systems, especially in distributed environments where events can be delayed due to network latency or processing lag. Without proper handling of late data, discrepancies can occur in the reconciliation process. The Watermark mechanism addresses this challenge by allowing a certain tolerance for late data. If an event arrives late, the watermark ensures that it is included in the processing sequence based on its actual event timestamp.

Mathematically, the late event handling process can be represented as:
\begin{equation}
    \text{Processed Time} = \max(\text{Event Time}, W(t)),
\end{equation}
where \( \text{Event Time} \) is the timestamp of the incoming event, and \( W(t) \) is the current watermark. If \( \text{Event Time} \) is greater than \( W(t) \), the event will be included in the processing window. This mechanism allows late events to be processed in the correct order, ensuring inventory reconciliation remains accurate despite delays in event arrival.

\subsubsection{Handling Out-of-Order Events}
In real-time systems, events often arrive out of order due to various factors such as network delays or concurrent operations at different terminals. The Watermark mechanism in Apache Flink solves this issue by providing a reliable way to handle out-of-order events. The core idea is to track the maximum event time and update the watermark to ensure that all events up to a certain point in time are processed, even if they are not received in the correct sequence. The mathematical representation of this can be written as:
\begin{equation}
    W(t) = \max(W(t-1), \text{Event Time}), \quad \forall \text{ events at } t
\end{equation}
where \( W(t) \) is the updated watermark, and \( \text{Event Time} \) represents the actual timestamp of an event. The system compares the event time with the current watermark and processes the event if it is within the valid window to ensure that out-of-order events are not discarded but rather processed correctly, maintaining inventory consistency.

\subsubsection{Improving Throughput and Real-Time Responsiveness}
One of the key advantages of using the Watermark mechanism in this project is its ability to improve system throughput while maintaining real-time responsiveness. Since Flink's Watermark mechanism ensures that events are processed based on event time, it helps maintain high throughput and low-latency processing without sacrificing data accuracy. The throughput of the system can be mathematically modeled as:
\begin{equation}
    \text{Throughput} = \frac{\text{Number of Events Processed}}{\text{Processing Time}}.
\end{equation}
By adjusting the maximum allowed lateness (\( \Delta t \)) in the watermark mechanism, we can balance between processing efficiency and event timeliness. The system can tolerate some level of delay (\( \Delta t \)) in event arrival, which allows it to process large volumes of data with minimal latency. This ensures that real-time inventory updates are synchronized with minimal delay, contributing to faster decision-making.

The application of the Watermark mechanism significantly enhances the reliability and accuracy of the inventory management system. It ensures that inventory events are processed in the correct order and that late or out-of-order events are included in the processing sequence, maintaining data consistency and preventing reconciliation errors. In mathematical terms, the Watermark ensures that:
\begin{equation}
    \text{Final Inventory State} = f(\text{All Event Times}, W(t)),
\end{equation}
where \( f \) represents the function that computes the final inventory state after processing all events. By controlling the event processing based on the Watermark mechanism, we guarantee that the final inventory state is consistent with the actual sequence of events, even in the presence of delayed or out-of-order data.

This capability improves the overall accuracy of inventory management, reduces manual intervention, and ensures real-time data synchronization between the warehouse system and the ERP system. By enabling the system to process large-scale data streams efficiently, the Watermark mechanism supports both high throughput and low latency, significantly enhancing decision-making processes and operational efficiency in real-time inventory reconciliation.

\subsection{ERP Synchronization and Fault Recovery}
Traditional ERP systems rely on centralized databases, making them vulnerable to data loss during network failures. To enhance fault tolerance, we implemented:
\begin{itemize}
    \item \textbf{Local Caching}: All stock transactions are temporarily stored on PDA devices, ensuring continuity during network failures.
    \item \textbf{Delta Synchronization}: When ERP connectivity is restored, the system automatically detects missing updates and synchronizes incremental changes.
    \item \textbf{Flink Push-Pull Mechanism}: Our system leverages Apache Flink’s dual \textbf{Push-Pull synchronization} mechanism. In the event of an ERP failure, Flink's \textbf{Pull Mode} ensures that inventory updates are stored in the event processing pipeline until the ERP connection is restored. Once back online, the \textbf{Push Mode} efficiently transmits accumulated data to the ERP system, maintaining consistency without data loss.
    \item \textbf{Watermark Mechanism}: Apache Flink's event-time processing ensures inventory records maintain strict temporal consistency, preventing data duplication or loss.
\end{itemize}

In this Flink Push-Pull Synchronization, a few modes available. In the \textbf{Push Mode}, under normal conditions, Flink continuously pushes real-time inventory updates to the ERP system, hence transactions are processed instantly if the ERP is operational. In the \textbf{Pull Mode}, if an ERP failure occurs, Flink buffers transaction data within a durable event processing pipeline, such as Apache Kafka or RocksDB. These transactions are timestamped and stored in the Flink state backend, ensuring no data is lost. Furthermore, in the \textbf{Automated Recovery}, once ERP services are restored, Flink automatically pulls and resends the buffered data in the correct chronological order, ensuring consistency between warehouse transactions and ERP records. Finally, a \textbf{Consistency Check} is performed before updating the ERP, Flink verifies that previously missed transactions are not duplicated, preventing inventory misalignment. This setup results in a recovery performance of 10,000 transactions local cache storage capacity, default 50,000 transactions of Flink event buffer size, and recovery speed less than 1 second upon ERP reconnection.

\subsection{Flutter-Based Mobile Terminal}
The system utilizes a Flutter-based mobile application for warehouse operators to perform stocktaking operations. The key functionalities of the mobile terminal include barcode scanning for quick and efficient inventory data collection, user-friendly interface designed for warehouse operators with minimal training requirements, and REST API communication with the backend server for real-time inventory updates. By leveraging a cross-platform mobile application framework, the system ensures compatibility with both Android and iOS devices, making deployment more flexible in different warehouse environments.

\section{EXPERIMENTS AND DISCUSSIONS}\label{sec:experimental_research}

\subsection{Experiment Setting}
To evaluate the performance of the proposed semi-automated inventory management system, we conducted experiments simulating diverse and complex warehouse scenarios. The experimental design incorporates realistic environments to ensure the applicability of results across different warehouse types. Additionally, a blind stocktaking test was implemented, where operators conducted inventory checks without prior knowledge of product locations, ensuring the authenticity of the results.

The experimental setup include: 500 inventory locations, 1,000 product batches, 37,000 storage units, and over 2 million individual products. We tested several key scenarios that include: high-density shelving (simulates tightly packed inventory to test readability under constrained spaces), dynamic inventory flow (models logistics centers with high product turnover), and medium to large electronics warehouse (with 20,000 square meters space).

Three groupings are introduced to test various scenarios (see \autoref{sec:a_exp_details}):
\begin{itemize}
    \item \textbf{Experimental Group:} Utilizes a Flutter-based stocktaking application, where operators use handheld terminals to scan and record product information. This system incorporates AI-driven anomaly detection and real-time inventory synchronization.
    \item \textbf{Control Group A (RFID System):} Uses traditional RFID technology, with fixed RFID readers scanning tagged products. No manual barcode scanning is involved, and stock discrepancies are resolved via ERP adjustments.
    \item \textbf{Control Group B (Manual Barcode Scanning):} Uses a conventional barcode scanner, requiring manual scanning of each item without AI-driven analytics or real-time anomaly detection.
\end{itemize}

To ensure fairness and consistency in all three systems, the only controlled difference between Control Group B (Manual Barcode Scanning) and Experimental Group (Flink + AI) is the integration of the Flink framework for real-time processing and AI-driven analysis for anomaly detection. All other conditions, including inventory scanning methods, user interface, data entry procedures, and operator workflows, were strictly identical to eliminate unintended biases.

\subsection{Experimental Procedure}
To ensure consistency and fairness across the three groups, the experiment was conducted in three distinct phases, detailed in \autoref{sec:a_exp_details}. The procedure aimed to evaluate the speed, accuracy, and usability of the inventory under controlled conditions. Furthermore, to ensure the objectivity and reliability of manual stocktaking, we implemented a double-blind review process. Two independent operators, Operator A and Operator B, conducted separate manual counts without prior knowledge of each other's results. The final manual count was established based on the following conditions to ensure that manual data was free from individual biases and served as a reliable benchmark for evaluating the proposed system's accuracy:
\begin{itemize}
    \item If the difference between Operator A and Operator B's manual counts was \(\leq 1\%\), their mean value was used as the golden standard.
    \item If the difference was 1\%, a supervisor conducted a third verification, and the majority agreement determined the final count.
\end{itemize}

\subsection{Participant Selection, Evaluation, and Training}
\label{sec:select_eval_train}

To ensure the reliability and consistency of the experiment, a rigorous selection process was conducted. A total of 40 candidates were initially selected based on the following criteria: at least 10 years of warehouse experience, familiarity with RFID inventory operations and manual recordkeeping, and prior stocktaking records with an accuracy of at least 95\%. To further ensure competence and minimize performance variations, all candidates were required to pass a pre-test in a simulated warehouse environment.  

Based on the results, only 13 participants who met all criteria were selected for the experiment. Furthermore, to ensure consistent performance across all groups, all operators underwent a standardized three-phase training program. This program covered the specific operations required for each experimental group and emphasized the importance of following predefined scanning protocols to ensure fairness and accuracy (see \autoref{sec:a_pretest_training} for details).

\subsection{Numerical Results and Analysis}

\subsubsection{Stocktaking Completion Time Analysis}

\begin{figure*}[htbp]
    \centering
    \includegraphics[width=0.7\textwidth]{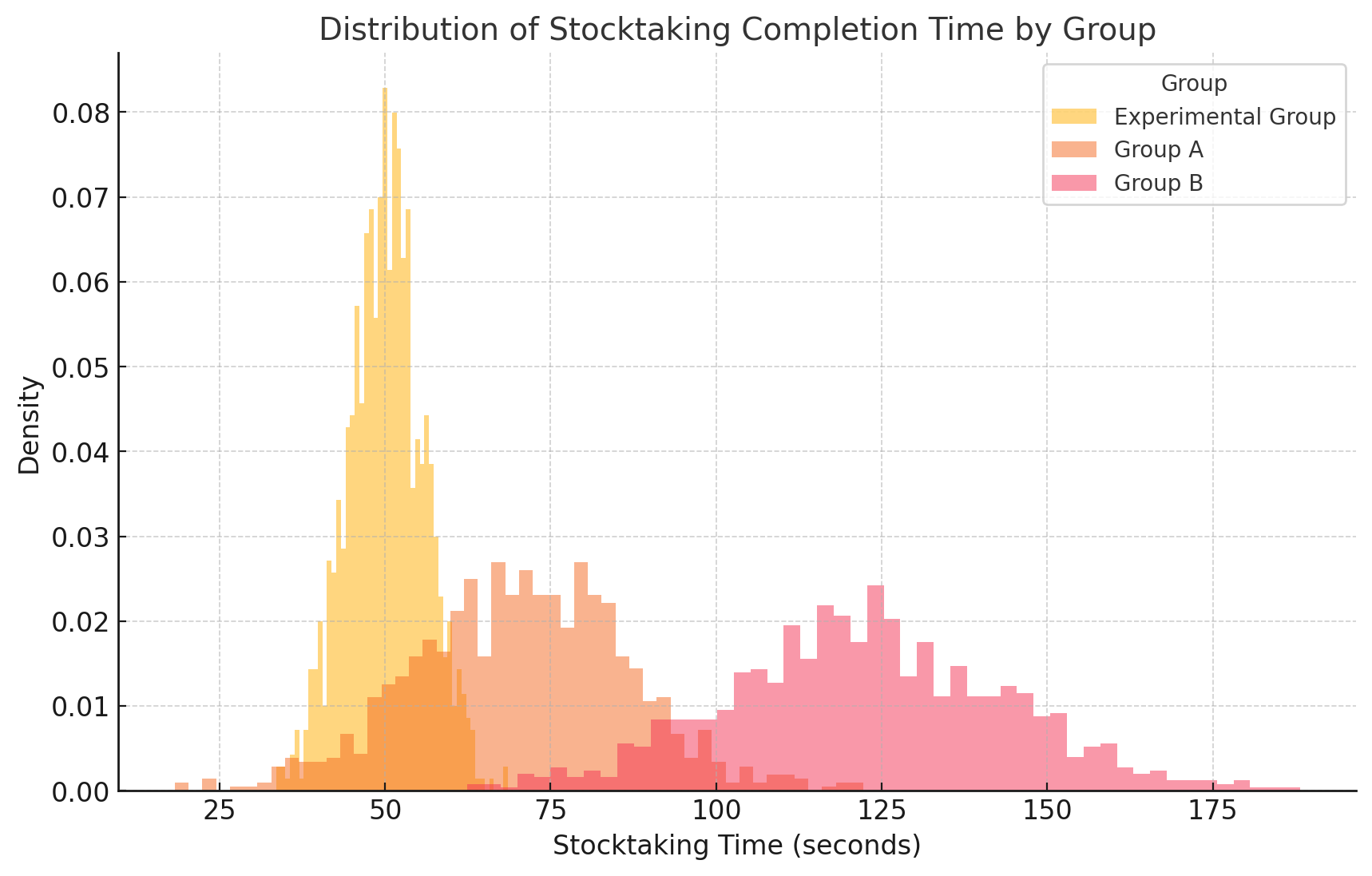}
    \caption{Stocktaking Completion Time Distribution}
    \label{fig:stocktaking-time-distribution}
\end{figure*}

The distribution of stocktaking completion time is presented in Figure~\ref{fig:stocktaking-time-distribution} with key statistics summarized in Table~\ref{tab:stocktaking_time}. The independent samples $t$-test highlights: 
\begin{itemize}
    \item Experimental Group vs. Group A: $t \approx -36.93$,
    \item Experimental Group vs. Group B: $t \approx -106.15$, and
    \item Control Group A vs. Group B: $t \approx -62.14$.
\end{itemize}
These all indicate that the corresponding $p$-values are far less than $0.001$. In addition, the One-Way ANOVA yielded an F statistic of approximately $5696$, demonstrating that the differences among the three groups are statistically significant ($p \ll 0.001$). Finally, for 1,000 sets of stocktaking data, the means and standard deviations for each set remain unchanged. Both the t-tests and the ANOVA confirm that the differences among the groups are extremely significant.

\begin{table}[htbp]
\sffamily \small
\centering
\caption{Stocktaking Completion Time with Sample Size}
\label{tab:stocktaking_time}
\footnotesize
\begin{tabular*}{\columnwidth}{lccc}
\toprule
\textbf{Group} & \textbf{Avg Time (s)} & \textbf{Std (s)} & \textbf{Samples (n)} \\
\midrule
Experimental Group & 50.37 & 5.60 & 1,000 \\
Group A & 70.76 & 16.53 & 1,000 \\
Group B & 123.21 & 20.97 & 1,000 \\
\bottomrule
\end{tabular*}
\normalsize\normalfont
\end{table}

\subsubsection{Inventory Accuracy and Error Analysis}
To quantify the accuracy of the proposed system, we use the mean squared error (MSE), root of mean squared error (RMSE), and the 95\%-confidence interval (CI) calculated as follows:
\begin{equation}
\text{MSE} = \frac{1}{N} \sum_{i=1}^{N} (\text{System Count}_i - \text{Manual Count}_i)^2,
\end{equation}
\begin{equation}
\text{RMSE} = \sqrt{\text{MSE}},
\end{equation}
\begin{equation}
\text{CI} = \bar{E} \pm 1.96 \times \frac{\sigma}{\sqrt{N}},
\end{equation}
where \( \text{System Count}_i \) is the stock count recorded by the system, \( \text{Manual Count}_i \) is the verified golden standard from double-blind manual stocktaking, \( \bar{E} \) is the mean error, and \( \sigma \) is the standard deviation of errors, and \( N \) is the number of stock items tested. The results are summarized in Table~\ref{tab:combined_inventory_analysis}.

\subsubsection{IF Evaluation}
To evaluate the IF model, we calculate the False Negative Rate (FNR) and False Positive Rate (FPR) as follows:
\begin{align}
    \text{FNR} &= \frac{\text{FN}}{\text{FN} + \text{TP}} \\
    \text{FPR} &= \frac{\text{FP}}{\text{FP} + \text{TN}},
\end{align}
where TP (True Positive) is the actual anomalies that were correctly detected, FN (False Negative) is the actual anomalies that were missed, FP (False Positive) is the false alarms (incorrectly flagged as anomalies), and TN (True Negative) is the inventory correctly classified as normal. The summary is reported in Table \ref{tab:combined_inventory_analysis}.

\begin{table*}[htbp]
\sffamily \small
\centering
\caption{Stocktaking Performance Comparison}
\label{tab:combined_inventory_analysis}
\footnotesize
\begin{tabular*}{\textwidth}{@{\extracolsep{\fill}}lcccccccccc}
\toprule
& \multicolumn{4}{c}{\textbf{Inventory Accuracy}} & \multicolumn{6}{c}{\textbf{IF Model}} \\
\cmidrule(lr){2-5} \cmidrule(lr){6-11}
\textbf{Group} & \textbf{Accuracy} & \textbf{MSE} & \textbf{RMSE} & \textbf{95\% CI} & \textbf{TP} & \textbf{TN} & \textbf{FP} & \textbf{FN} & \textbf{FPR (\%)} & \textbf{FNR (\%)} \\
\midrule
Experimental Group (Flutter + AI) & 98.5\% & 0.85 & 0.92 & (0.79, 0.94) & 28 & 957 & 13 & 2 & 1.3\% & 6.7\% \\
Group A (RFID) & 97.0\% & 1.73 & 1.32 & (1.61, 1.85) & 60 & 910 & 20 & 10 & 2.2\% & 14.3\% \\
Group B (Manual Barcode) & 95.0\% & 2.45 & 1.56 & (2.21, 2.68) & 75 & 875 & 45 & 5 & 4.9\% & 6.3\% \\
\bottomrule
\end{tabular*}
\normalsize\normalfont
\end{table*}

To determine the optimal parameters for the IF model, we conducted a parameter tuning experiment using Grid Search. The experiment varied the following hyperparameters: number of estimators \{100, 200, 500\}, max samples per tree \{128, 256, 512\}, contamination rate \{0.005, 0.01, 0.02, 0.05\}, and max features per split \{3, 5, 7\}. The optimized hyperparameter results are shown in \autoref{sec:a_hyperparam_opt}. Based on the results, we selected contamination = 0.009, as it provides the best balance between low FNR (1.3\%) and FPR (1.8\%), while achieving a minimal RMSE of 0.92. The results show that the AI-assisted Flutter system is faster and more accurate than both RFID and manual stocktaking methods. The use of IF for anomaly detection also provided significant improvements in anomaly detection with minimal false positives or negatives. These findings are important as they show the potential of combining AI and mobile technologies for streamlining inventory management, making it faster, more accurate, and cost-effective compared to traditional systems.

\subsubsection{Participants' Experimental Feedback}
Structured interviews were conducted to gather qualitative feedback. To ensure sample diversity, participants included operators with varying experience levels and usage frequencies (from occasional to frequent). The interviews focused on five key domains, with specific questions designed for each domain. The questions for each domain and groups feedback are available in \autoref{sec:interview}.

\subsubsection{Cost Comparison Across Scenarios}
To assess the costs associated with the \textbf{full RFID implementation}, we estimated system acquisition and implementation expenses using data from our experimental setup. We assume that RFID readers will require periodic replacement due to technological advancements and wear-and-tear. Additionally, integration with enterprise resource planning (ERP) systems (e.g., SAP, IBM) significantly contributes to the cost.

For RFID to be fully effective, all inventory items must be tagged during implementation. If items are not tagged, inventory inaccuracy will persist, leading to increased reliance on manual stocktaking. Using 30 seconds per tag, we calculated the total labor cost required for initial tagging based on an assumed workforce wage. Finally, to determine the initial RFID tagging cost, we also estimated the average cost per tag based on a 90/10 ratio (normal vs. metal tags). During implementation, the total estimated cost of tags and labor is \$3,700. These cost breakdowns are detailed in \autoref{sec:a_full_rfid}.

For the \textbf{semi-automated barcode implementation}, the costs were significantly lower than RFID. The cost of barcode tags is significantly lower than RFID. Based on an average printed barcode tag cost of \$0.04, the implementation cost for barcode tagging, including labor. The cost breakdowns for the semi-automated system are detailed in \autoref{sec:a_semi_automated}.

We calculated the return-on-investment (ROI) period based on setup and maintenance costs, as shown in Table~\ref{tab:roi-comparison}. From a cost perspective, the barcode system has significantly lower initial investment and maintenance costs, making it more suitable for businesses with limited budgets or those that do not require high-efficiency management in the short term. However, the RFID system has clear advantages in efficiency and automation, as it can quickly scan multiple items, does not require direct line-of-sight, and is resistant to contamination. It is ideal for complex supply chains and environments requiring high precision. While the initial investment for an RFID system is higher, its long-term operational efficiency and inventory management accuracy offer greater value. Overall, the RFID system is more suited for businesses that require efficient, real-time inventory management, while the barcode system is better for small businesses or organizations with limited resources that have lower efficiency demands.

\begin{table}[htbp]
\sffamily \small
\centering
\caption{ROI Comparison}
\label{tab:roi-comparison}
\footnotesize
\begin{tabular*}{\columnwidth}{lcc}
\toprule
\textbf{Cost Type} & \textbf{Semi-Automated System} & \textbf{RFID System} \\
\midrule
Initial Setup Cost & \$16,830 & \$545,150 \\
Maintenance & Low & Medium to High \\
ROI Period & 6–12 months & 24–36 months \\
\bottomrule
\end{tabular*}
\normalsize\normalfont
\end{table}

\section{CONCLUSIONS}\label{sec:conclusions}
Next research will explore enhancements such as a combination of drone scanning with stereo camera and digital twins with federated learning for cross-warehouse predictions, and AR-assisted scanning, providing a scalable and progressive path for intelligent inventory management digitalization without disrupting existing operations. Warehouse inventory management faces persistent challenges with fully automated RFID technology, such as high implementation costs, sensitivity to environmental interference, and network load imbalances, driving the need for a more practical and cost-effective solution to enhance efficiency and accuracy in modern supply chains. This study proposes a semi-automated inventory management system that integrates barcode scanning, real-time data stream processing, and AI-driven analytics to address these issues. Experimental validation demonstrates that the system achieves a 98.5\% inventory accuracy in complex warehouse scenarios, surpassing traditional RFID systems by 1.5\%, while reducing equipment costs to just 3\% of RFID solutions. By leveraging Apache Flink’s watermarking mechanism and Isolation Forest anomaly detection, it cuts inventory synchronization time from 45 minutes to under 5 seconds and reduces manual verification workload by 76\%, offering a lightweight architecture that seamlessly connects mobile terminals, stream processing, and ERP systems. The system’s cost-effectiveness, with an 82\% lower deployment cost than RFID, and its adaptability to challenging environments like metal-rich or cold chain storage, make it an ideal solution for medium to large warehouse facilities with daily turnovers below 500,000 items.

\section*{ACKNOWLEDGEMENTS}
The author acknowledges the support of warehouse staff and technical teams who participated in the experimental validation of the proposed system. No specific funding sources were utilized for this research.

\section*{CONFLICTS OF INTEREST}
The author declares no conflicts of interest.

\section*{DATA AVAILABILITY STATEMENTS}
Due to the nature of the research, commercial supporting data is not available.

\appendix
\newpage

\section*{APPENDICES}
\section{Experiment Details}
\label{sec:a_exp_details}

The conditions are summarized in Table~\ref{tab:experimental-conditions}. The procedures are described in Table~\ref{tab:exp_procedure_phase}.

\begin{table*}[h]
\sffamily \small
\centering
\caption{Experimental Conditions Across Groups}
\label{tab:scenario_group_test}
\footnotesize
\begin{tabular*}{\textwidth}{@{\extracolsep{\fill}}p{0.3\textwidth}p{0.7\textwidth}}
\toprule
\textbf{Condition} & \textbf{Description} \\
\midrule
\textbf{Tag Placement and Orientation} & For RFID (Control Group A): RFID tags were placed uniformly across all items, with reader distances controlled (e.g., 0.5m, 1m, and 2m). For Manual Barcode (Control Group B) and Flutter+Barcode (Experimental Group): Barcode labels were uniformly placed on items. \\
\textbf{Obstacle Simulation} & Metal Shelves: Standard warehouse-grade metal racks with a height of 2 meters, spaced 0.5 meters apart. These were used to simulate potential RFID signal distortion and to evaluate barcode scanning accessibility in tight spaces. Stacked Cardboard Boxes: Cardboard stacks with a height of 1.5 meters, placed to create blind spots. For RFID, they assessed signal penetration through non-metal materials; for barcode systems, they measured the operator's ability to access and scan items in less visible areas. \\
\textbf{Environmental Conditions} & All scenarios were conducted under standard indoor lighting conditions of approximately 500–700 lux and a consistent room temperature of 22°C ± 2°C to ensure uniform visibility and operability. \\
\textbf{Operator Standardization} & Operators across all groups were trained using a three-phase training process (see Table \ref{tab:operator_training}) to minimize performance variability. \\
\textbf{Sample Size} & 500 inventory locations, 1,000 product batches, and 37,000 storage units were tested in all groups. Products were selected using stratified sampling to ensure representativeness across diverse categories. \\
\bottomrule
\end{tabular*}
\normalsize\normalfont
\label{tab:experimental-conditions}
\end{table*}

\begin{table*}[h]
\sffamily \small
\centering
\caption{Experimental Procedure Phases}
\label{tab:exp_procedure_phase}
\footnotesize
\begin{tabular*}{\textwidth}{@{\extracolsep{\fill}}p{0.25\textwidth}p{0.7\textwidth}}
\toprule
\textbf{Phase} & \textbf{Description} \\
\midrule
\textbf{Preparation Phase} & Products were labeled: \begin{itemize} \item RFID tags for \textbf{Control Group A}. \item Barcode labels for \textbf{Control Group B} and \textbf{Experimental Group}. \end{itemize} Warehouse environment was standardized: \begin{itemize} \item Identical obstacle placement for all groups (see Table~\ref{tab:experimental-conditions}). \item Consistent lighting and temperature conditions. \end{itemize} Operators completed \textbf{standardized training} to ensure proficiency. \\
\textbf{Execution Phase} & \textbf{Control Group A (RFID System)}: \begin{itemize} \item Operators used an \textbf{RFID scanner} at a \textbf{90° angle}, swinging their arm like a radar while maintaining a \textbf{1-meter} distance. \item The system automatically \textbf{logged scanned batches}; any unscanned batch triggered an \textbf{anomaly alert}. \item After stocktaking, operators \textbf{manually reviewed} unscanned batches and performed a \textbf{second scan}. \item If the second scan failed, the batch was marked as \textbf{potentially missing}, and an \textbf{exception report} was submitted. \end{itemize} \textbf{Control Group B (Manual Barcode)}: \begin{itemize} \item Operators manually scanned \textbf{each barcode}, ensuring \textbf{line-of-sight visibility}. \item Data was entered into \textbf{Microsoft Excel} using a \textbf{PDA device}. \item Any missed items were \textbf{manually recorded} on paper for later review. \end{itemize} \textbf{Experimental Group (Flutter + Barcode + AI)}: \begin{itemize} \item Operators used \textbf{PDA terminals} for inventory stocktaking. \item The \textbf{Isolation Forest anomaly detection algorithm} calculated inventory anomalies. \item Any batch with an \textbf{anomaly score} \( S_{\text{total}} \) above the \textbf{95\% confidence threshold} was flagged for \textbf{manual review}. \item Operators reviewed flagged batches: \begin{itemize} \item \textbf{Operator A:} Rechecked flagged stocktaking data, identified missing scans, and corrected the data. \item \textbf{Operator B:} Verified anomalies; if a batch was misplaced, they manually corrected the count and informed the warehouse manager. \end{itemize} \end{itemize} \\
\textbf{ERP Cross-Validation} & \begin{itemize} \item If \textbf{discrepancies remained} between system inventory and ERP records, a \textbf{final review} was conducted. \item If necessary, the warehouse team \textbf{manually adjusted} stock data in the ERP system. \item The system confirmed the final stock levels and flagged any unresolved issues. \end{itemize} \\
\bottomrule
\end{tabular*}
\normalsize\normalfont
\label{tab:experimental_procedure}
\end{table*}

\section{Pre-Test Criteria and Training Phases}
\label{sec:a_pretest_training}

The pre-test criteria for participants are summarized in Table~\ref{tab:exp_pre_test_eval}. The training phases are summarized in Table~\ref{tab:operator_training}.

\begin{table*}[h]
\sffamily \small
\centering
\caption{Pre-Test Evaluation Criteria}
\label{tab:exp_pre_test_eval}
\footnotesize
\begin{tabular*}{\textwidth}{@{\extracolsep{\fill}}p{0.3\textwidth}p{0.6\textwidth}}
\toprule
\textbf{Evaluation Criterion} & \textbf{Requirement} \\
\midrule
Stocktaking Accuracy & Achieving \textbf{\(\geq 98\%\)} accuracy in inventory counts. \\
Stocktaking Speed Consistency & Maintaining a standard deviation of stocktaking time within \textbf{15\%}. \\
Error Handling & Demonstrating the ability to identify and correct stock discrepancies. \\
\bottomrule
\end{tabular*}
\normalsize\normalfont
\label{tab:pretest_criteria}
\end{table*}

\begin{table*}[h]
\sffamily \small
\centering
\caption{Operator Training Phases}
\label{tab:operator_training}
\footnotesize
\begin{tabular*}{\textwidth}{@{\extracolsep{\fill}}p{0.3\textwidth}p{0.5\textwidth}}
\toprule
\textbf{Training Phase} & \textbf{Details} \\
\midrule
\textbf{Basic Training (2 hours)} & Introduction to system-specific features, including: \begin{itemize} \item For Experimental Group (Flutter + Barcode): Training covered system navigation, barcode scanning, error handling, and real-time anomaly detection using the Flutter app. \item For Control Group A (RFID System): Operators were trained in both radar scanning (90° fixed angle) and freehand scanning methods. Radar scanning, which achieved a higher recognition rate of approximately 95\%, was emphasized to ensure consistency, while limitations of freehand scanning (recognition rate ~70\%) were noted. \item For Control Group B (Manual Barcode): Training focused on handheld barcode scanning and database entry procedures. \end{itemize} \\
\textbf{Practical Training (6 hours)} & Operators conducted simulated stocktaking trials under controlled conditions: \begin{itemize} \item For RFID systems, operators practiced maintaining the optimal 90° scanning angle to ensure consistent recognition rates. \item For barcode systems, operators practiced achieving accurate line-of-sight scans for various product placements. \end{itemize} \\
\textbf{Assessment Test (1 hour)} & Operators completed 10 rounds of simulated stocktaking: \begin{itemize} \item Error rates were monitored and kept below 10\%. \item Standard deviation of stocktaking times was controlled at $<5\%$. \item Operators with error rates exceeding thresholds underwent additional training to ensure convergence. \end{itemize} \\
\bottomrule
\end{tabular*}
\normalsize\normalfont
\label{tab:operator-training}
\end{table*}

\section{Hyperparameter Optimization}
\label{sec:a_hyperparam_opt}

For the IF model, the hyperparameters are optimized, with results summarized in Table~\ref{tab:hyperparameter-results}.

\begin{table*}[htbp]
\sffamily \small
\centering
\caption{Hyperparameter Optimization Results}
\label{tab:hyperparameter-results}
\footnotesize
\begin{tabular*}{\textwidth}{@{\extracolsep{\fill}}cccc}
\toprule
\textbf{Contamination} & \textbf{FNR (\%)} & \textbf{FPR (\%)} & \textbf{RMSE} \\
\midrule
0.005 & 1.1\% & 2.5\% & 0.98 \\
0.009 & 1.3\% & 1.8\% & 0.92 \\
0.01  & 1.6\% & 1.5\% & 1.05 \\
0.02  & 2.4\% & 1.2\% & 1.32 \\
0.05  & 4.8\% & 0.8\% & 2.45 \\
\bottomrule
\end{tabular*}
\normalsize\normalfont
\end{table*}

\section{Interview Question and Results}\label{sec:interview}

The questions are summarized in Table~\ref{tab:interview-questions}. The qualitative feedback is summarized in Table~\ref{tab:qualitative-feedback}. The key metrics across different groups are summarized in Table~\ref{tab:group-comparison}.

\begin{table*}[h!]
\sffamily \small
\centering
\caption{Structured Interview Questions for Qualitative Feedback}
\footnotesize
\begin{tabular*}{\textwidth}{@{\extracolsep{\fill}}p{5cm}p{10cm}}
\toprule
\textbf{Domain} & \textbf{Interview Questions} \\
\midrule
\textbf{Usability} & 1. What do you find most convenient or challenging about using the system? \newline 2. How would you describe the system's learning curve? Does it require extensive training? \newline 3. How easy was it to get started with the system initially? Were there sufficient guidelines? \newline 4. Have you encountered any unclear operational logic while using the system? \\
\addlinespace[1ex]
\textbf{Comfort and Ergonomics} & 1. How comfortable is the system during prolonged use? Do you experience fatigue or discomfort? \newline 2. Does the weight and design of the scanning device affect your experience? If so, how? \newline 3. Do you typically use the system while standing or sitting? Does the device accommodate your posture? \newline 4. Is the layout of the system's hardware (e.g., screen, buttons) convenient to operate? \\
\addlinespace[1ex]
\textbf{Work Efficiency} & 1. How does the system impact your work efficiency? What are the improvements or shortcomings compared to traditional methods? \newline 2. How long do you use the system daily? How do you feel about it? \newline 3. Does the system help you complete repetitive tasks faster? In what ways? \newline 4. During busy periods, does the system effectively support your work needs? \\
\addlinespace[1ex]
\textbf{Interface and Interaction Design} & 1. What is your opinion on the intuitiveness of the system's interface? Is it easy to navigate? \newline 2. How is the system's response speed in high-intensity scenarios? Do you experience lag or delays? \newline 3. Are the icons, colors, and text in the interface clear and easy to understand? \newline 4. Do you feel that certain interaction steps could be simplified? \\
\addlinespace[1ex]
\textbf{Improvement Suggestions} & 1. What suggestions do you have for improving the interface or workflow? \newline 2. What features would you like to see added to enhance work efficiency or comfort? \newline 3. In your work context, is there any support that the system lacks? If so, what? \newline 4. Which aspects of the hardware or software do you think need the most optimization? \\
\bottomrule
\end{tabular*}
\normalsize\normalfont
\label{tab:interview-questions}
\end{table*}

\begin{table*}[h!]
\sffamily \small
\centering
\caption{Summary of Qualitative Feedback from Operator Interviews}
\label{tab:qualitative-feedback}
\footnotesize
\begin{tabular*}{\textwidth}{@{\extracolsep{\fill}}p{2.5cm}p{3.5cm}p{3.5cm}p{3.5cm}}
\toprule
\textbf{Domain} & \textbf{Positive Feedback} & \textbf{Issues and Suggestions} & \textbf{Key Statistics} \\
\midrule
\textbf{Usability} & 80\% of operators found the system intuitive and easy to learn & 30\% suggested simplifying the interface for high-intensity use; 15\% noted multitasking difficulties & 80\% became proficient within 1-2 days \\
\addlinespace[1ex]
\textbf{Ergonomics} & 76\% reduction in manual validation workload & 30\% found the scanner heavy after 4+ hours; 20\% suggested wearable options & 76\% workload reduction \\
\addlinespace[1ex]
\textbf{Efficiency} & 90\% reported significant efficiency gains; 70\% noted fewer errors & 25\% experienced delays during peak hours & 90\% efficiency improvement \\
\addlinespace[1ex]
\textbf{Interface Design} & 75\% praised the intuitive design and navigation & 40\% suggested better visual feedback; 35\% requested voice commands & 75\% UI satisfaction \\
\addlinespace[1ex]
\textbf{Improvement Suggestions} & N/A & Interface: improve feedback, add voice commands; Hardware: lighter scanners, flexible mounting; Features: personalization, automation & N/A \\
\bottomrule
\end{tabular*}
\normalsize\normalfont
\end{table*}

\begin{table*}[h!]
\sffamily \small
\centering
\caption{Comparison of Workload, Usability, and Satisfaction Across Experimental Groups}
\label{tab:group-comparison}
\footnotesize
\begin{tabular*}{\textwidth}{@{\extracolsep{\fill}}p{0.3\textwidth}p{0.2\textwidth}p{0.2\textwidth}p{0.2\textwidth}}
\toprule
\textbf{Dimension} & \textbf{Control Group A} & \textbf{Control Group B} & \textbf{Experimental Group} \\
\midrule
\textbf{Workload} & Requires tag installation; time-consuming & Requires scanning each barcode manually; moderate workload & Only requires scanning barcodes; AI-assisted process reduces effort \\
\textbf{Installation Complexity} & Requires tag adhesion and testing & Simple barcode labeling & Simple barcode labeling, no additional setup required \\
\textbf{Instant Feedback} & Device debugging is complex & No real-time analytics, only raw scan data & Immediate feedback with AI-driven anomaly detection \\
\textbf{Repetitive Labor} & High; may lead to fatigue & High; scanning required for every item & Low; AI minimizes repetitive actions \\
\textbf{Ease of Use} & Rating: 3.5/5 & Rating: 4.2/5 & Rating: 4.8/5 \\
\textbf{Training Time} & Requires 6+ hours & Requires 3+ hours & Mastered in 2 hours \\
\textbf{Satisfaction} & Rating: 3.2/5 & Rating: 3.9/5 & Rating: 4.7/5 \\
\bottomrule
\end{tabular*}
\normalsize\normalfont
\end{table*}

\section{Costs for Full RFID System}
\label{sec:a_full_rfid}

For the full RFID implementation, the breakdown for system cost, labor cost, and initial cost is shown in Table~\ref{tab:rfid-system-cost}, Table~\ref{tab:rfid-labor-cost}, and Table~\ref{tab:rfid-tag-sensitivity}.

\begin{table*}[h!]
\sffamily \small
\centering
\caption{Estimated RFID System Acquisition Costs}
\label{tab:rfid-system-cost}
\footnotesize
\begin{tabular*}{\textwidth}{@{\extracolsep{\fill}}lccc}
\toprule
\textbf{Item} & \textbf{Quantity} & \textbf{Unit Cost} & \textbf{Total Cost} \\
\midrule
Fixed Readers & Variable & \$500--\$1,000 & \$500--\$1,000 per unit \\
RFID Tags & 37,000 & \$0.20 & \$7,400 \\
Middleware Integration & 1 license & Varies & \$500,000 \\
Tag and Reader Testing & 1 test & Estimated & \$1,000 \\
Software Maintenance (Annual) & 1 & Estimated & \$36,000 \\
\textbf{Total System Acquisition Cost} & — & — & \textbf{\$545,150} \\
\bottomrule
\end{tabular*}
\normalsize\normalfont
\end{table*}

\begin{table*}[htbp]
\sffamily \small
\centering
\caption{Estimated Labor Costs of RFID Implementation}
\label{tab:rfid-labor-cost}
\footnotesize
\begin{tabular*}{\textwidth}{@{\extracolsep{\fill}}lcccc}
\toprule
\textbf{\# Storage Units} & \textbf{Minutes/Tag} & \textbf{Total Hours} & \textbf{Wage/Hour} & \textbf{Total Labor Cost} \\
\midrule
37,000 & 0.6 & 370 & \$10 & \$3,700 \\
\bottomrule
\end{tabular*}
\normalsize\normalfont
\end{table*}

\begin{table*}[htbp]
\sffamily \small
\centering
\caption{Sensitivity Analysis of Initial RFID Tag Cost}
\label{tab:rfid-tag-sensitivity}
\footnotesize
\begin{tabular*}{\textwidth}{@{\extracolsep{\fill}}lccccc}
\toprule
\textbf{Tag Type} & \textbf{Avg. Cost/Tag} & \textbf{10/90} & \textbf{25/75} & \textbf{50/50} & \textbf{75/25} \\
\midrule
Normal Item & \$0.10 & \$370 & \$925 & \$1,850 & \$2,775 \\
Metal Item & \$1.50 & \$49,950 & \$41,625 & \$27,750 & \$13,875 \\
\textbf{Total Cost} & — & \$50,320 & \$42,550 & \$29,600 & \$16,650 \\
\bottomrule
\end{tabular*}
\normalsize\normalfont
\end{table*}

\section{Costs for Semi-Automated System}
\label{sec:a_semi_automated}

The system cost for the semi-automated system is summarized in Table~\ref{tab:barcode-system-cost}. The tagging cost is summarized in Table~\ref{tab:barcode-tag-cost}.

\begin{table*}[htbp]
\sffamily \small
\centering
\caption{Estimated Semi-Automated System Acquisition Costs}
\label{tab:barcode-system-cost}
\footnotesize
\begin{tabular*}{\textwidth}{@{\extracolsep{\fill}}lccc}
\toprule
\textbf{Item} & \textbf{Quantity} & \textbf{Unit Cost} & \textbf{Total Cost} \\
\midrule
Barcode Scanning Devices & 11 & \$300 & \$3,300 \\
Software/Middleware & 1 & Included & Included \\
\textbf{Total System Acquisition Cost} & — & — & \textbf{\$3,300} \\
\bottomrule
\end{tabular*}
\normalsize\normalfont
\end{table*}

\begin{table*}[htbp]
\sffamily \small
\centering
\caption{Tagging Costs at Implementation}
\label{tab:barcode-tag-cost}
\footnotesize
\begin{tabular*}{\textwidth}{@{\extracolsep{\fill}}lccccc}
\toprule
\textbf{\# Storage Units} & \textbf{Minutes/Tag} & \textbf{Total Hours} & \textbf{Wage/Hour} & \textbf{Tag Price/Tag} & \textbf{Total Cost} \\
\midrule
37,000 & 0.083 & 51.39 & \$39.06 & \$0.04 & \$13,530 \\
\bottomrule
\end{tabular*}
\normalsize\normalfont
\end{table*}

\end{document}